\documentclass[aps,preprint,showpacs]{revtex4}
\usepackage{graphicx,latexsym,amsmath}
\begin{document}
\title{Helical polymer in
cylindrical confining geometries}
\author{
A. Lamura$^{1,2}$, T. W. Burkhardt$^{2,3}$, and G. Gompper$^{2}$ }
\affiliation{$^1$ Istituto Applicazioni Calcolo, CNR, Sezione di
Bari,
via Amendola 122/D, 70126 Bari, Italy \\
$^2$ Institut f\"ur Festk\"orperforschung, Forschungszentrum
J\"ulich, D-52425 J\"ulich, Germany \\
$^3$ Department of Physics, Temple University, Philadelphia, PA
19122, USA}
\begin{abstract}
Using an algorithm for simulating equilibrium configurations, we study 
a fluctuating
helical polymer either (i) contained in a cylindrical pore or (ii) wound around a
cylindrical rod. We work in the regime where both the contour length and the
persistence length of the helical polymer are  much larger than the diameter of the
cylinder. In case (i) we calculate the free energy of confinement and interpret it
in terms of a worm-like chain in a pore with an effective diameter that depends on
the parameters of the helix. In case (ii) we consider the possibility that one end
of the helical polymer escapes from the rod and wanders away. The average numbers of
turns at which the helix escapes or intersects the rod are measured in the
simulations, as a function of the pitch $p_0$. The behavior for large and small
$p_0$ is explained with simple scaling arguments.
\end{abstract}

\pacs{87.15.Aa, 36.20.Ey, 61.25.Hq, 82.70.-y} 
\maketitle

\section{Introduction}
In this paper we study some of the equilibrium statistical properties of a confined
helical or ribbon-like polymer. The cases of (i) a polymer contained in a
cylindrical pore and (ii) a polymer wound around a cylindrical rod are considered.
Some motivation is provided by the following observations:

Biological polymers differ from synthetic polymers in that they are semi-flexible,
with a persistence length much larger than the monomer size, and usually have a
helical structure. This is well known for DNA, but F-actin also has a double-helical
structure, while microtubuli are helical cylinders. The diameter of these biological
polymers is in the range of 1 to 25 nm. Polymeric helical structures are also found
in self-assembling systems, consisting of either amphiphiles or peptides. In some
cases the diameters and pitch lengths are much larger than for the biopolymers
mentioned above.

In amphiphilic systems the formation of helical ribbons has been observed in
multicomponent mixtures of a bile salt or some other nonionic surfactant,
phosphatidylcholine or a fatty acid, and a steroid analog of cholesterol
\cite{chun93,zast99}. The ribbons have typical diameters in the range of 5 to 20
$\mu$m, and pitch angles between 10 and $60^o$. Other examples are ethanolic/water
solutions of diacetylenic phospholipids, in which the formation of hollow tubules of
diameter 0.6 $\mu$m and typical lengths of 10 to 100 $\mu$m has been observed
\cite{yage84,schn93,thom99,pakh03}. Helically-coiled phospholipid-bilayer ribbons
appear as metastable intermediates in the growth of these tubules.

Other systems, which show spontaneous assembly of ribbons, are aqueous solutions of
peptides \cite{zhan93,agge97,nyrk00a,nyrk00b}. Depending on the solution conditions,
the same peptide exists in different conformations, such as random coils,
$\alpha$-helices, or $\beta$-sheets. At not too low peptide concentrations, the
molecules self-assemble into long $\beta$-sheet structures which form twisted
ribbons (with a straight central axis). The width of these ribbons is about 4 nm,
and their length is of the order of 500 nm \cite{nyrk00a,nyrk00b}. These ribbons can
aggregate due to face-to-face attraction into twisted fibrils of a thickness of 8-10
nm.

Interestingly, in a self-assembling system of gemini surfactants (two surfactant
molecules covalently linked at their charged head group), the degree of twist and
the pitch of the micrometer-scale ribbons has been found to be tunable by the
introduction of opposite-handed chiral counterions \cite{oda99}.

The confinement of polymers in cylindrical tubes is one of the classical problems in
polymer physics. For biological polymers, such a confinement occurs, for example,
when viral DNA of a bacteriophage squeezes through the narrow tail during DNA
injection.  Technological advances in the manipulation of single molecules in micro-
and nanofluidic devices \cite{chou00,gior01} has fueled interest in the structure
and dynamics of biological polymers in confined geometries \cite{jend03}.

Helical and twisted ribbons can be confined not only by external walls, but also by
winding around each other, as in the fibril formation of twisted $\beta$-sheet
peptides mentioned above. The simple model we consider, consisting of a helical
wound around a thin cylinder, is a step in this direction but leaves out some
important physical features, such as the face-to-face attraction in the fibrils.

\section{Free Energy of Confinement}
The free energy $\Delta F$ of confinement of a fluctuating polymer
of contour length $\ell$ in a cylindrical pore of diameter $D$ is
defined by
\begin{equation}
\exp(-\Delta F/k_BT)=\frac{Z(D,\ell)}
{Z(\infty,\ell)}=p(\ell)\;\label{freeenergy}
\end{equation}
Here $Z(D,\ell)$ and $Z(\infty,\ell)$ are the partition functions
of the polymer with one end fixed in the presence and absence of
the cylindrical confining geometry, respectively. The quantity
$\Delta F$ represents the work required to squeeze the polymer
reversibly into the cylindrical pore. It may be evaluated in
simulations by generating polymer configurations with one fixed
end in an infinite volume with the Boltzmann probability,
computing the fraction $p(\ell)$ of the configurations of arc
length $\ell$ which lie entirely within a cylindrical domain of
diameter $D$, and making use of Eq.~(\ref{freeenergy})

For a flexible, self-avoiding polymer with vanishing bending rigidity, $\Delta F$ is
purely entropic. The confinement of such a polymer in a cylindrical pore is
considered in Refs. \cite{degennes,bg,ps}.

In the worm-like chain model of a semi-flexible polymer, the
polymer is represented by an inextensible line or filament ${\bf
r}(s)$ with contour length $\ell$ and elastic energy
\begin{equation}
E_{\rm worm}=\frac{\kappa}{2}\int_0^\ell\left(\frac{d{\bf t}_3}
{ds}\right)^2\thinspace ds\;.\label{wormlike}
\end{equation}
Here $s$ specifies distance along the contour, ${\bf t}_3=d{\bf
r}/ds$ is the unit tangent vector, $\kappa$ is the bending
rigidity, and $P=\kappa/k_BT$ is the the persistence length. In
the narrow-pore, long-polymer limit $D\ll P\ll \ell$, the polymer
is almost a straight line, i.e. the angle between the tangent
vector ${\bf t}_3$ and the $z$ axis or symmetry axis of the
cylinder is a small quantity. In this case the right-hand side of
Eq.~(\ref{freeenergy}) decays as
\begin{equation}
p(\ell)\sim e^{-E_0\ell}\;\label{pell}
\end{equation}
for large $\ell$, where $\exp(-E_0\thinspace dz)$ is the largest
eigenvalue of the transfer matrix of a slice of the system of
thickness $dz$. The quantity $E_0^{-1}$ represents a typical
contour length at which the configurations intersect the pore
wall. According to Eqs.~(\ref{freeenergy}) and (\ref{pell}) the
confinement free energy per unit length $\Delta f=\Delta F/\ell$
is given by
\begin{equation}
\frac{\Delta f}{k_BT}=E_0(P,D)=\frac{A_\circ}
{P^{1/3}D^{2/3}}\;,\label{circle}
\end{equation}
where the dependence on $P$ and $D$ follows from simple scaling or
dimensional arguments \cite{odijk,dijkstra,twb97}. Similarly, for
a pore with a rectangular cross section with edges $L_1,L_2\ll P$,
\begin{equation}
\frac{\Delta f}{k_BT}=E_0(P,L_1,L_2)=\frac{A_\Box}
{P^{1/3}}\left(\frac{1}{L_1^{2/3}}+\frac{1}
{L_2^{2/3}}\right)\;.\label{rectangle}
\end{equation}
The quantities $A_\circ$ and $A_\Box$ on the right-hand sides of
Eqs.~(\ref{circle}) and (\ref{rectangle}) are dimensionless
universal numbers $A_\circ$ and $A_\Box$, which are the same for
all worm-like chains.

The prediction $A_\Box=1.1036$ was obtained in Ref. \cite{twb97} by solving an
integral equation numerically which arises in an exact analytical approach.
Measuring the probability $p(\ell)$ in Eq.~(\ref{pell}) in simulations, fitting the large $\ell$ behavior with the exponential form (\ref{pell}), 
and making use of Eqs.~(\ref{circle}) and (\ref{rectangle}), 
Bicout and Burkhardt \cite{Bicout} estimated
\begin{equation}
A_\circ=2.375\pm0.013\;,\quad\quad A_\Box=1.108\pm
0.013\;.\label{bicout}
\end{equation}
An earlier estimate from simulations, $A_\circ=2.46\pm0.07$, was
given by Dijkstra {\it et al.} \cite{dijkstra}.

\section{Helical Polymer Model}

In this paper we generalize the above results to helical polymers
or chiral ribbons, which have spontaneous curvature and torsion.
Again the polymer is replaced by a curve ${\bf r}(s)$ of fixed
contour length ${\cal S}$. To each point on the line a
right-handed triad of unit vectors ${\bf t}_1(s), {\bf
t}_2(s),{\bf t}_3(s)$ is assigned, where ${\bf t}_3=d{\bf r}/ds$
is the tangent vector and ${\bf t}_1$, ${\bf t}_2$ correspond to
principal axes of the polymer cross section. The rotation of the
triad along the curve is governed by the generalized Frenet
equations \cite{PanRabPRL00,PanRabPRE00,KKRabPRE02}
\begin{equation}
\frac{d {\bf t}_i}{d s}={\boldsymbol\omega}\times{\bf t}_i\;,\quad
{\boldsymbol\omega}={\bf t}_1\omega_1+{\bf t}_2\omega_2+{\bf
t}_3\omega_3\;,\label{frenet1}
\end{equation}
or
\begin{equation} \frac{d {\bf
t}_i}{ds}=\sum_{j,k}\epsilon_{ijk}{\bf
t}_j\omega_k\;.\label{frenet2}
\end{equation}
The elastic energy is given by
\cite{PanRabPRL00,PanRabPRE00,KKRabPRE02}
\begin{equation}
E_{\rm helix} = \frac{1}{2} \sum_{j=1}^{3} b_j \int_0^{\cal S} ds
\left[\omega_j(s) - \omega_{0j}(s)\right]^2\;, \label{energy}
\end{equation}
where the coefficient $b_1$ and $b_2$ are bending rigidities along
the principal axes of the cross section, and $b_3$ is the twist
rigidity. The parameters $\omega_j(s)$ and $\omega_{0j}(s)$
determine the curvatures and torsions in the deformed and
stress-free states of the polymer, respectively. Since the energy
is quadratic in the deviations $\delta \omega_j =
\omega_j-\omega_{0j}$, the distribution of $\delta \omega_j$ is
Gaussian, with zero mean and second moment
\begin{equation}
\langle\delta \omega_i(s) \delta \omega_j(s^{'})\rangle =
\frac{k_BT}{b_i}\thinspace\delta_{ij}\thinspace\delta(s -
s^{'})\;.\label{moments}
\end{equation}

We restrict our attention to the case $\omega_{0j}(s)={\rm
constant}$, corresponding to a helical polymer with spontaneous
curvature and torsion but without spontaneous twist. In the
absence of fluctuations, i.e. in the limit $b_1=b_2=b_3=\infty$,
the Frenet equations are readily solved \cite{explain1}, yielding
\begin{eqnarray}
{\bf r}(s)={\bf r}(0)&+&\frac{1}{\omega_0}\Biggl\{{\bf
t}_3(0)\sin(\omega_0s)+{\bf e}(0)\thinspace\omega_{03}\left[s-
\frac{\sin(\omega_0 s)}{\omega_0}\right]
\nonumber\\
&+& {\bf e}(0)\times{\bf t}_3(0)\left[1-\cos(\omega_0
s)\right]\Biggr\}\;,\label{nofluc1}
\end{eqnarray}
where
\begin{equation}
{\bf e}(s)={\bf t}_1(s)\frac{\omega_{01}}{\omega_0}+{\bf
t}_2(s)\frac{\omega_{02}}{\omega_0}+ {\bf
t}_3(s)\frac{\omega_{03}}{\omega_0}\;, \quad
\omega_0=\left(\omega_{01}^2+\omega_{02}^2+\omega_{03}^2\right)^{1/2}\;.\label{nofluc2}
\end{equation}
Equation (\ref{nofluc1}) represents a helix with radius $r_0$ and
pitch $p_0$, where
\begin{equation}
r_0=\frac{(\omega_{01}^2+\omega_{02}^2)^{1/2}}{\omega_0^2}\;,\quad
p_0=2\pi\frac{\omega_{03}}{\omega_0^2}\;,\label{radiuspitch}
\end{equation}
winding around an axis pointing in the direction of the unit
vector ${\bf e}(0)$.

Including Gaussian fluctuations according to Eq.~(\ref{moments}),
Panyukov and Rabin \cite{PanRabPRL00,PanRabPRE00} showed that
\begin{equation}
\langle{\bf t}_i(s)\cdot{\bf t}_j(0)\rangle=\left(e^{-{\bf
\Gamma}\thinspace s}\right)_{ij}\;, \end{equation} where
${\bf\Gamma}$ is the matrix with elements
\begin{equation}
\Gamma_{ij}= \frac{1}{2}k_BT
\left(\sum_{k}b_k^{-1}-b_i^{-1}\right)\delta_{ij}-
\sum_k\epsilon_{ijk}\omega_{0k}\;.
\end{equation}
The two-point correlation function of the unit vector ${\bf e}(s)$
in Eq.~(\ref{nofluc2}), which is directed along the axis of the
helix, follows from this result. In the special case
$b=b_1=b_2=b_3$ considered in our simulations,
\begin{equation}
\langle{\bf e}(s)\cdot{\bf e}(0)\rangle=e^{-s/L_p}\;, \quad
L_p=\frac{b}{k_BT}\;,\label{expdecay}
\end{equation}
where $L_p$ is the persistence length.

\section{Simulations}

Following Kats {\it et al.} \cite{KKRabPRE02}, we replace the
differential equations (\ref{frenet2}) by the difference equations
\begin{equation}
t_{i k}(s+ds)=\sum_{j}O_{ij}t_{j k}(s)\label{differenceeq}
\end{equation}
in our simulations. Here $t_{i k}$ denotes the $k$-th component of ${\bf t}_i$ with
respect to a fixed Cartesian coordinate system, $O$ is the orthogonal matrix
\begin{equation}
O=\left(1+\frac{1}{2}A\thinspace ds\right)\left(1-\frac{1}{2}
A\thinspace ds\right)^{-1}\;,\label{matrixO}
\end{equation}
and $A$ is the antisymmetric matrix with elements
$A_{ij}=\sum_k\epsilon_{ijk}\omega_k$. The difference equations
are consistent with the Frenet equations (\ref{frenet2}) to first
order in $ds$, and the orthogonality of the matrix $O$ preserves
the orthonormality of the ${\bf t}_i$ in the simulations.

For simplicity we set $b=b_1=b_2=b_3$, corresponding to 
Eq.~(\ref{expdecay}). In accordance with Eq.~(\ref{moments}), 
the $\delta\omega_j(s)$ are chosen randomly
from a Gaussian distribution with zero mean and standard deviation
$(k_BT/b\thinspace ds)^{1/2}$, where $ds\ll L_p=b/k_BT$.

\section{Helical polymer in a cylindrical pore}

We have determined the confinement free energy of a helical
polymer fluctuating in a narrow cylindrical pore from simulations.
Cylinders with both circular and square cross sections were
considered, and we use the same symbol $D$ for the diameter and
edge, respectively. The symmetry axis of the cylinder defines the
$z$ axis of our fixed coordinate system.

The helical polymer was generated step by step using the numerical procedure
described in the preceding Section. The radius $r_0$ and pitch $p_0$ were set to
desired values by choosing $\omega_{01}$, $\omega_{03}$ in 
Eq.~(\ref{radiuspitch}) appropriately, with $\omega_{02}=0$. The 
starting point was chosen randomly, apart
from the requirements that the stress-free helix fit inside the cylinder, with its
axis parallel to the $z$ axis. This is the case for the initial vectors ${\bf
t}_1(0) = (\omega_{03}/\omega_0,0,\omega_{01}/\omega_0)$, ${\bf t}_2(0) = (0,1,0)$,
and ${\bf t}_3(0) = (-\omega_{01}/\omega_0,0,\omega_{03}/\omega_0)$, which were
used. The other simulation parameters were $ds=10^{-4}$, $D=1$, $L_p=b_1/k_B T =
b_2/k_B T = b_3/k_B T = 8000$. Clearly $ds \ll D \ll L_p\thinspace$. The pore is
narrow in comparison with the persistence length, and $ds$ is small in order to
approximate the continuum model (\ref{frenet1}).

To obtain the free energy of confinement of the helical polymer,
we proceeded as discussed below Eq.~(\ref{freeenergy}), generating
many polymer configurations and computing the probability $P(n)$
that the polymer has not yet intersected the pore wall
\cite{explain2} after $n$ steps of the algorithm. The
determination of $P(n)$ was based on $50,000$ independent helices.
For large $n$ an exponential decay
\begin{equation}
P(n)\sim e^{-\lambda_0 n}\;,\label{expdecay2}
\end{equation}
similar to the result (\ref{pell}) for semi-flexible polymers, is
expected. According to Eq.~(\ref{freeenergy}), the free energy of
confinement per unit length along the axis of the helix is given
by
\begin{equation}
\frac{\Delta f}{k_BT}=\frac{\lambda_0}{\xi ds}\;,\quad
\xi=\frac{p_0}{\ \left[p_0^2+(2\pi
r_0)^2\right]^{1/2}}\;.\label{freeenergy2}
\end{equation}
Here we use the relation $\ell=\xi s$ between the contour length
$s$ and the corresponding length $\ell$ along the axis of the
helix. The persistence length $L_p$, defined with respect to the
contour length as in Eq.~(\ref{expdecay}), and the corresponding
persistence length $P$, defined with respect to length along the
axis of the helix, also satisfy $P=\xi L_p$.

A helical polymer with persistence length $L_p$ in a pore with diameter $D\ll L_p$
has the same confinement free energy as a semi-flexible polymer with persistence
length $P=\xi L_p$ in a pore with effective diameter $D_{\rm eff}$. To define
$D_{\rm eff}$ quantitatively, we equate the free energies of confinement
(\ref{circle}) and (\ref{freeenergy2}), obtaining
\begin{equation}
\frac{A_\circ}{D_{\rm eff}^{2/3}}=\frac{(\xi L_p)^{1/3}}{\xi
ds}\lambda_0\;,\quad \frac{A_\Box}{D_{\rm eff}^{2/3}}=
\frac{(\xi L_p)^{1/3}}{2\xi ds}\lambda_0\;.\label{Deff}
\end{equation}

The probability $P(n)$ for $r_0=p_0=0.3$, with the other
simulation parameters noted above, is shown in Fig. 1. The data
are in good agreement with the exponential decay
(\ref{expdecay2}), and the values of $\lambda_0$ are given in the
figure caption. As in the case of a semi-flexible polymer
\cite{Bicout}, the curves $P(n)$ for the circular and square
cylinders practically coincide when plotted versus $\lambda_0 n$
instead of $n$.

For the exponential decay (\ref{expdecay2}), the mean number of steps of the
algorithm at which the polymer intersects the wall equals $\lambda_0^{-1}$,
corresponding to $N_i=\xi ds/(p_0\lambda_0)$ turns of the helix. The values of
$\lambda_0$ in the caption of Fig. 1 yield $N_i=8.0$ and $8.7$ for the circular and
square cross sections. Since the number of turns before intersecting the wall is
fairly large, the helix should be equivalent to a worm-like chain in a pore of width
$D_{\rm eff}=D - 2 r_0$. To check the equivalence quantitatively, we use 
Eq.~(\ref{Deff}) with $D_{\rm eff}=D - 2 r_0$ and the values of 
$\lambda_0$ in the
caption of Fig. 1 to predict the amplitudes $A_\circ$, $A_\Box$. This yields
\begin{equation}
A_{\circ} = 2.45 \pm 0.05\;,\quad\quad A_{\Box} = 1.12 \pm 0.04\;,
\end{equation}
in good agreement with the results (\ref{bicout}) for
semi-flexible polymers.

We have also studied the dependence of $D_{\rm eff}$ on the
polymer pitch $p_0$, keeping the radius $r_0$ and the persistence
length $L_p$ constant. For small $p_0$ the polymer makes many
turns before intersecting the wall and is equivalent to a
semi-flexible polymer in a pore of diameter $D-2r_0$, as discussed
in the preceding paragraph. In the limit $p_0 \to\infty$, the
helical polymer does not make any turns before intersecting the
wall and corresponds to a semi-flexible polymer in a pore of
diameter $D$. As $p_0$ increases from $0$ to $\infty$, $D_{\rm
eff}$ is expected to vary monotonically between these two limiting
values.

For various values of $p_0$ we have computed the probability $P(n)$ 
that a helical polymer with radius $r_0=0.3$, persistence length 
$L_p=8000$, and contour length $nds$ in a cylindrical pore with a 
circular cross section of diameter $D=1$ does not
intersect the wall. The corresponding $\lambda_0$ was obtained from 
an exponential fit (\ref{expdecay2}) for large $n$. Finally 
$D_{\rm eff}$ was calculated using Eq.~(\ref{Deff}) and the best 
estimate (\ref{bicout}) for $A_\circ$. The results are
shown in Fig.~2. The data do indeed interpolate between the expected 
limiting values $D-2r_0=0.4$ and $D=1$ for small and large $p_0$, 
respectively.

The crossover region in Fig.~2, where $D_{\rm eff}$ varies most rapidly 
with $p_0$, is around $p_0\approx 40$, $D_{\rm eff}\approx 0.7$. 
According to Eqs.~(\ref{pell}), (\ref{circle}), and (\ref{freeenergy2}), 
these values of $p_0$ and $D_{\rm eff}$ correspond to 
$N_i=(\xi L_p)^{1/3}D_{\rm eff}^{2/3}/(A_\circ p_0)\approx 0.2$ turns
of the helix before intersecting the wall.

\section{Helical polymer encircling a cylindrical rod}

In this Section we consider a helical polymer wound around a long cylindrical rod
with a circular cross section and diameter $D\ll L_p$. We study the possibility that
the polymer generated in the simulation escapes from the rod as $n$ increases and
wanders away.

In the simulations the parameters $ds=10^{-4}$, $b_1/k_B T = b_2/k_B T = b_3/k_B T =
L_p=8000$ were the same as in the preceding section. The diameter of the rod was
$D=0.2$, and the radius of the helix was $r_0=0.3$. For these parameters $ds\ll
D<2r_0\ll L_p\thinspace$. The starting point of the polymer was chosen randomly,
apart from the requirements that the stress-free helix wind around the cylindrical
rod without touching it, with the axis of the helix parallel to the rod.

>From the simulation data we computed the probability $P(n)$ that after $n$ steps the
polymer has not yet intersected the rod \cite{explain2}. Each curve $P(n)$ is based
on $10,000$ independent helices. The results for three different values of the pitch
$p_0$ are shown in Fig.~3. Unlike the case of a polymer in a cylindrical pore, shown
in Fig. 1, $P(n)$ does not decay to zero as $n$ increases. Instead, above a
characteristic value which depends on the pitch, the curve flattens and approaches a
nonzero limiting value. This is because the polymer generated in the simulation
sometimes escapes from the rod, due to a sufficiently large fluctuation, and wanders
away as $n$ increases, without ever returning to intersect the rod.

A simple theory of the escape, which suggests $P(n)=A+Be^{-C n}$,
in qualitative consistency with Fig. 3, is given in the Appendix.

We determined the average number of turns at which the helix escapes from the rod or
intersects it by making two checks after each step of the growth algorithm: (i) If
the distance of the endpoint ${\bf r}(s)$ from the axis of the rod is less than
$D/2$, the polymer has intersected the rod. (ii) If the distance of the endpoint
${\bf r}_{\rm axis}(s)$ of the {\em axis} of the helix is greater than $r_0+D/2$,
the circular cross section of the helix no longer encircles the rod, i.e. the helix
has escaped. Geometrically ${\bf r}_{\rm axis}(s)$ is determined as follows: Since
the unit vectors ${\bf t}_3(s)$ and ${\bf e}(s)$ are tangent to the helix and
directed along its axis, respectively, ${\bf e}(s)\times{\bf t}_3(s)$ is directed
perpendicularly from the point ${\bf r}(s)$ on the helix contour toward the
corresponding point ${\bf r}_{\rm axis}(s)$ on the axis of the helix. Thus,
\begin{eqnarray}
{\bf r}_{\rm axis}(s)&=&{\bf r}(s)+r_0\thinspace \frac{{\bf
e}(s)\times {\bf t}_3(s)}{|{\bf e}(s)\times{\bf t}_3(s)|}\nonumber\\
&=& {\bf r}(s)+{\bf t}_1(s) \frac{\omega_{02}}{\omega_0^2}-{\bf
t}_2(s) \frac{\omega_{01}}{\omega_0^2}\;,\label{raxis}
\end{eqnarray}
where we have used Eqs.~(\ref{nofluc2}) and (\ref{radiuspitch}).

In Fig. 4 the average numbers of turns $N_e^1$, $N_e^2$ at which the helix escapes
from the rod and the average number of turns $N_i$ at which the helix intersects the
rod are shown as functions of $p_0$. For each value of $p_0$, 10,000 independent
configurations were generated. Each configuration was continued until it intersected
the rod \cite{explain2} or the number of steps of the algorithm exceeded $5\times
10^6$, whichever came first. The average $N_e^1$ is based on all configurations
which escape, independent of whether they return to intersect the rod or not. The
quantity $N_e^2$ is the average value for only those configurations which escape and
in $5\times 10^6$ steps of the algorithm still have not intersected the rod.

For $p_0\leq 1$ the probability of the polymer escaping from the rod is so small
that $N_e^{1,2}$ could not be determined reliably with configurations of $5\times
10^6$ steps. For larger $p_0$, the data for $N_e^1$ and $N_e^2$ practically
coincide, indicating that the polymer rarely returns to intersect the cylinder once
it has escaped. The data are in excellent agreement with $N_e^{1,2},N_i\sim
p_0^{-1}$ for large $p_0$ and $N_i\sim p_0^{-2/3}$ for small $p_0$. These power laws
may be understood as follows.

The transverse fluctuations of the endpoint of the axis of an {\em unconfined}
helical polymer of length $\ell$ and persistence length $P$ about 
the corresponding endpoint of the unstressed helix are readily 
calculated from Eq.~(\ref{expdecay}) and given by \cite{explain3}
\begin{equation}
\langle r_\perp^2\rangle=\frac{2}{3}\thinspace \frac{\ell^3}{P}
\label{fluc}
\end{equation}
for $\ell\ll P$. Here both $\ell=\xi s$ and $P=\xi L_p$ are
measured along the axis of the helix, as discussed below 
Eq.~(\ref{freeenergy2}). Equation (\ref{fluc}) also applies to the
worm-like chain. Apart from the factor $2/3$, Eq.~(\ref{fluc}) 
follows from simple scaling or dimensional arguments
\cite{odijk,dijkstra,twb97}. The powers of $r_\perp$, $\ell$, and
$P$ in Eq.~(\ref{fluc}) are also consistent with $ E_0 \ell\sim 1$
and $r_\perp\sim D$ in Eqs.~(\ref{pell}) and (\ref{circle}).

Replacing $r_\perp$ in Eq.~(\ref{fluc}) by $r_0+D/2=0.4$, as in the
simulations, and solving for $N_e=\ell/p_0$ yields the estimate
\begin{equation}
N_e=\frac{12}{p_0^{2/3}\left(p_0^2+3.6\right)^{1/6}}\;.\label{Ne}
\end{equation}
for the number of turns of the helix at which the typical transverse displacement
equals the value needed for escape from the rod. Roughly speaking, the polymer
wrapped around the rod escapes in $N_e$ turns, as given by Eq.~(\ref{Ne}) for
$N_e\stackrel{<}{\sim} 1$, i.e. $p_0\stackrel{>}{\sim} 12$. For smaller $p_0$, the
typical transverse fluctuations in a single turn of the helix are too small for
escape from the rod, and the helix is more likely to intersect the rod than to
escape. Equation (\ref{Ne}), which ignores this possibility, no longer applies. As
noted above, for $p_0<1$ the escape probability is too small for a reliable
determination of $N_e^{1,2}$ with configurations of $5\times 10^6$ steps.

In the region $N_e\stackrel{<}{\sim} 1$, i.e. $p_0\stackrel{>}{\sim} 12$, 
Eq.~(\ref{Ne}), which corresponds to the solid curve in Fig. 4, is in 
good quantitative agreement with the simulation data for $N_e^{1,2}$. 
For large $p_0$, $N_e\approx 12/p_0$. The coefficient 12 is an 
order-of-magnitude estimate that happens to give a
good fit to the simulation data, whereas the power law 
$N_e\sim p_0^{-1}$ for large $p_0$ is exact. An argument based on 
Eq.~(\ref{fluc}) similar to the one for $N_e$
predicts $N_i\sim p_0^{-1}$ for large $p_0$, in excellent agreement with Fig. 4.
Note that the data points for $N_e^{1,2}$ and $N_i$ practically coincide.

Since the possibility of escape is negligible for small $p_0$, the helical polymer
is equivalent to a worm-like chain in a pore with diameter $D_{\rm eff}=2r_0-D$. To
estimate the average number of turns $N_i$ at which the polymer intersects the rod,
we replace $D$ by $D_{\rm eff}$ in Eqs.~(\ref{pell}) and (\ref{circle}) and solve
for the typical intersection length $\ell\approx E_0^{-1}$. This yields $\ell\sim
P^{1/3}$. Substituting  $r_\perp\sim D_{\rm eff}$ in Eq.~(\ref{fluc}) and solving
for $\ell$ leads to the same result. According to the discussion below 
Eq.~(\ref{freeenergy2}), $P=\xi L_p\approx L_p(p_0/2\pi r_0)$ for $p_0\ll r_0$. Keeping
track of the powers of $p_0$, we obtain the power law $N_i=\ell/p_0\sim p_0^{-2/3}$
for small $p_0$, in excellent agreement with the simulation data in Fig. 4

In Fig. 5 the fractions $f_e^1$, $f_e^2$, and $f_i$ of the 10,000 configurations
which contribute to $N_e^1$, $N_e^2$, and $N_i$ in Fig. 4 are shown as functions of
$p_0$. For $p_0\stackrel{<}{\sim}1$, $f_e^{1,2}\approx 0$ and $f_i\approx 1$, i.e.
almost all the configurations intersect the cylinder and never escape. Around
$p_0\approx 10$ the curves cross, and for larger $p_0$ the polymer is more likely to
escape than to intersect the rod.

\section{Concluding Remarks}

We have studied some statistical properties of a helical polymer in cylindrical
restrictive geometries of diameter $D$, in the limit that the persistence length $P$
along the axis of the helix is large in comparison with $D$ and the radius $r_0$ of
the helix. In this limit the helical polymer has much in common with the worm-like
chain. We interpret the simulation data for the free energy of confinement in a
cylindrical pore using the scaling form (\ref{circle}) for a worm-like chain in a
pore, with an effective diameter $D_{\rm eff}$ that is renormalized by the helical
structure. As the pitch $p_0$ of the helix increases from $0$ to $\infty$, $D_{\rm
eff}$ increases monotonically from $D-2r_0$ to $D$, as shown in Fig. 2.

Thinking in terms of a worm-like chain also proves useful in connection with the
escape of the helical polymer encircling a cylindrical rod. In the limit $P\gg r_0$
the transverse fluctuations of the axis of the helix are given by the same result
(\ref{fluc}) as for the worm-like chain. As $p_0$ increases, the typical transverse
displacement in one turn of the helix also increases, resulting in a greater
probability per turn of escape. We have used Eq.~(\ref{fluc}) to estimate the
average number of turns at which the helix escapes from the rod or intersects it.
The simulation data in Fig. 4 are in excellent agreement with the predicted
asymptotic forms $N_e,N_i\sim p_0^{-1}$, $N_i\sim p_0^{-2/3}$ for large and small
$p_0$, respectively.

It would be interesting to include an attractive interaction between the rod and the
polymer wound around it. In the fibril formation mentioned in the Introduction, the
attraction is an essential ingredient.

\acknowledgments AL and TWB thank Gerhard Gompper and coworkers
for hospitality at the Forschungszentrum J\"ulich. GG acknowledges 
the hospitality of the Aspen Center for Physics during the final 
stages of this work.

\appendix
\section{Simple theory of escape of a polymer}
Let us define $P^{ne}_N$ as the probability that the polymer has
neither intersected the cylindrical surface of the rod nor escaped
in the first $N$ turns of the helix and $P^e_N$ as the probability
that it has not yet intersected the cylindrical surface but that
that it has escaped. Treating each turn of the helix as
statistically independent, we denote the probability that a
polymer which has not yet escaped does escape in the next turn by
q, and the probability that it neither escapes nor intersects the
rod in the next turn by p. (The third possibility, that it
intersects the rod in the next turn, has probability 1-q-p.) In
addition we assume that once the polymer escapes, it never
intersects the rod.

These assumptions imply the recurrence relations
\begin{eqnarray}
&&P^{ne}_{N+1}=pP^{ne}_N\;,\label{recurr1}\\
&&P^e_{N+1}=qP^{ne}_N+P^e_N\;,\label{recurr2}
\end{eqnarray}
with initial conditions $P^{ne}_0=1$, $P^e_0=0$. Writing down the
first few iterates, it is easy to see that
\begin{eqnarray}
&&P^{ne}_N=p^N\;,\label{Pne}\\
&&P^e_N=q\thinspace \frac{1-p^N}{1-p}\;.\label{Pe}
\end{eqnarray}
 The probability
 $P_N=P^{ne}_N+P^e_N$ that the polymer has not yet intersected the
 rod after $N$ steps is analogous to $P(n)$ in Section
 V. From Eqs.~(\ref{Pne}) and (\ref{Pe})
 \begin{equation}
P_N=\frac{q}{1-p}+\frac{1-p-q}{1-p}\thinspace p^N\;.\label{decay}
\end{equation}
Thus, as $N$ increases, $P_N$ decays exponentially from $P_0=1$ to
$P_\infty=q/(1-p)$. The mean number of turns $N_e$ at which escape occurs is given
by
\begin{equation}
N_e=\frac{\sum_{N=1}^\infty N[P^e_N-P^e_{N-1}]}{\sum_{N=1}^\infty
[P^e_N-P^e_{N-1}]}=\frac{1}{1-p}\;.
\end{equation}
An analogous calculation for the mean number of turns $N_i$ at which the polymer
insects the rod yields $N_i=N_e$.

This theory is obviously an oversimplification, but the form (\ref{decay}) of the
decay, $P_N=A+Be^{-C N}$, is qualitatively consistent with Fig. 3. \vfill\eject

\newpage

\begin{figure}[ht]
\begin{center}
\includegraphics*[width=0.7\textwidth]{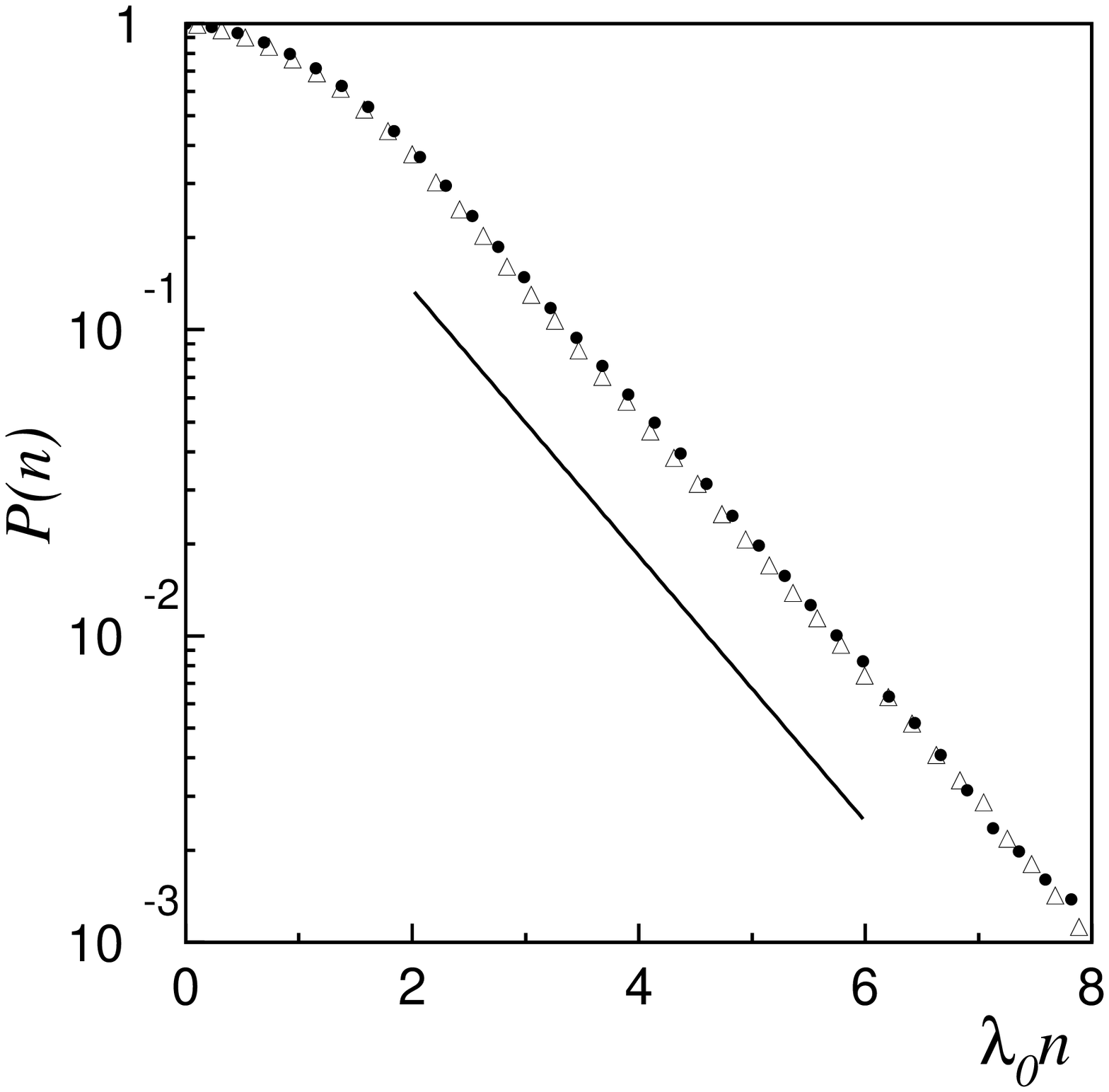}
\end{center}
\caption{Probability $P(n)$ that a helical polymer with radius $r_0=0.3$, pitch
$p_0=0.3$, and persistence length $L_p=8000$ in a cylindrical pore does not
intersect the pore wall in the first $n$ steps of the algorithm. The 
full circles ($\bullet$) correspond to a pore with a circular cross 
section with diameter $D=1$ and the triangles ($\triangle$) to a 
square cross section with edge length $D=1$. Fitting the data to 
Eq.~(\ref{expdecay2}) for large $n$ yields $\lambda_0=6.57\times
10^{-6}$ and $6.01\times 10^{-6}$, respectively. The full line 
corresponds to the
exact exponential decay $e^{-\lambda_0 n}$.} \label{fig1}
\end{figure}

\clearpage
\newpage

\begin{figure}[ht]
\begin{center}
   \includegraphics*[width=0.7\textwidth]{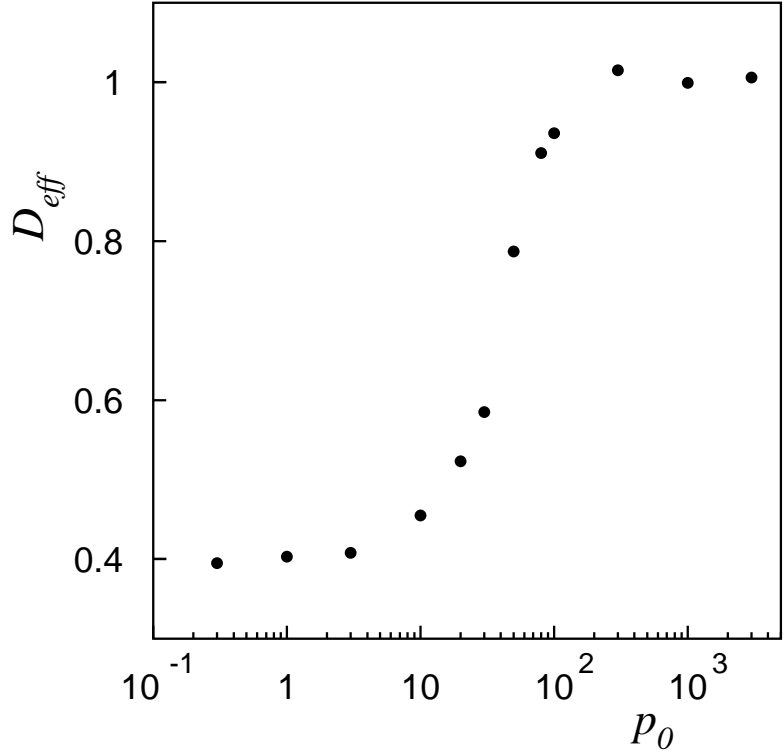}
\end{center}
\caption{The effective diameter $D_{\rm eff}$ as a function of the
pitch $p_0$ for a helical polymer with radius $r_0=0.3$ and
persistence length $L_p=8000$ in a cylindrical pore with a
circular cross section with diameter $D=1$. The data interpolate
between the limiting values $D-2r_0=0.4$ and $D=1$ for small and
large $p_0$, respectively.} \label{fig2}
\end{figure}

\clearpage
\newpage

\begin{figure}[ht]
\begin{center}
   \includegraphics*[width=0.7\textwidth]{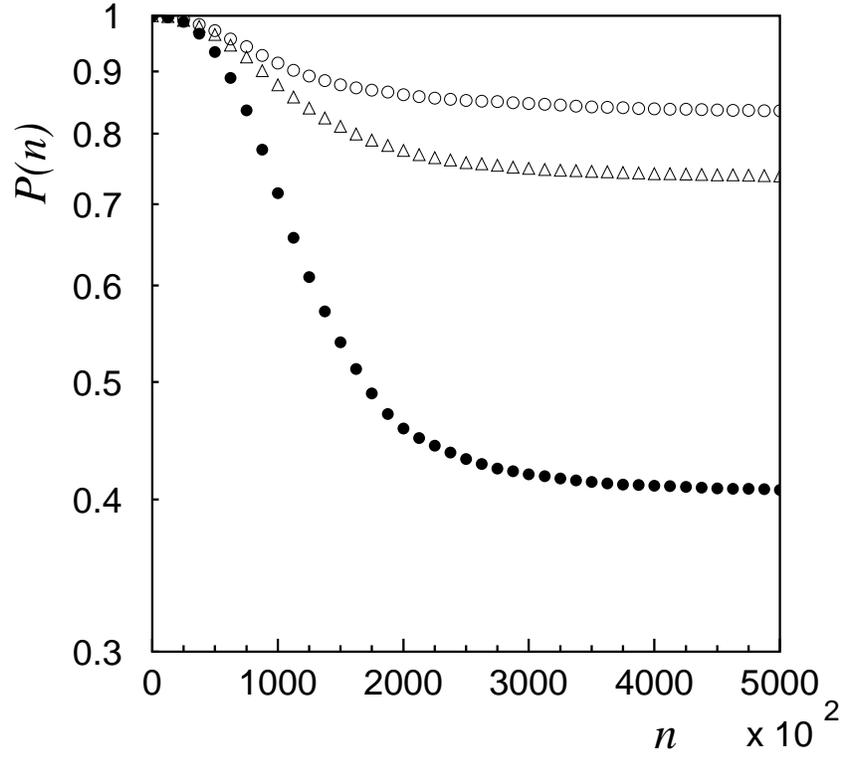}
\end{center}
\caption{Probability $P(n)$ that a helical polymer with radius $r_0=0.3$ and
persistence length $L_p=8000$, wound at the fixed end around a cylindrical rod of
diameter $D=0.2$ does not intersect the rod in the first $n$ steps of the algorithm.
The pitch of the helix is $p_0=10$ ($\bullet$), 30 ($\triangle$), and 100
($\circ$).} \label{fig3}
\end{figure}

\clearpage
\newpage

\begin{figure}[ht]
\begin{center}
   \includegraphics*[width=0.7\textwidth]{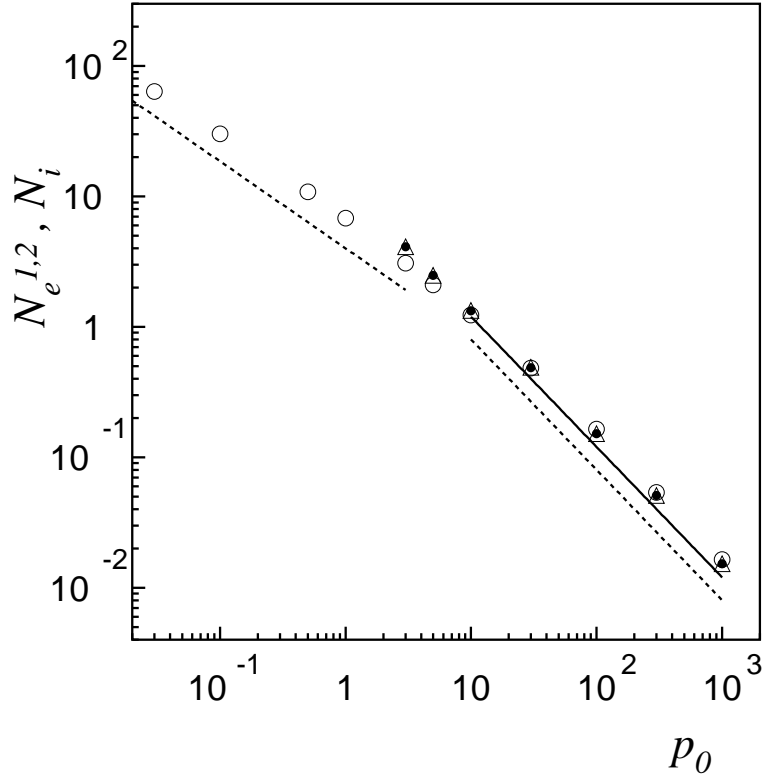}
\end{center}
\caption{Average numbers of turns $N_e^1$ ($\triangle$), $N_e^2$ ($\bullet$) at
which the helix escapes from the rod, and the average number of turns $N_i$
($\circ$) at which the helix intersects the rod as a function of $p_0$. Here $N_e^1$
is based on all the configurations which escape, independent of whether they return
to intersect the rod or not; $N_e^2$ is based on the configurations which escape and
in $5\times 10^6$ steps of the algorithm do not return to intersect the rod. The
dashed lines on the left and right have slopes -2/3 and -1, respectively, and the
solid line shows the prediction (\ref{Ne}). } \label{fig4}
\end{figure}

\clearpage
\newpage

\begin{figure}[ht]
\begin{center}
   \includegraphics*[width=0.7\textwidth]{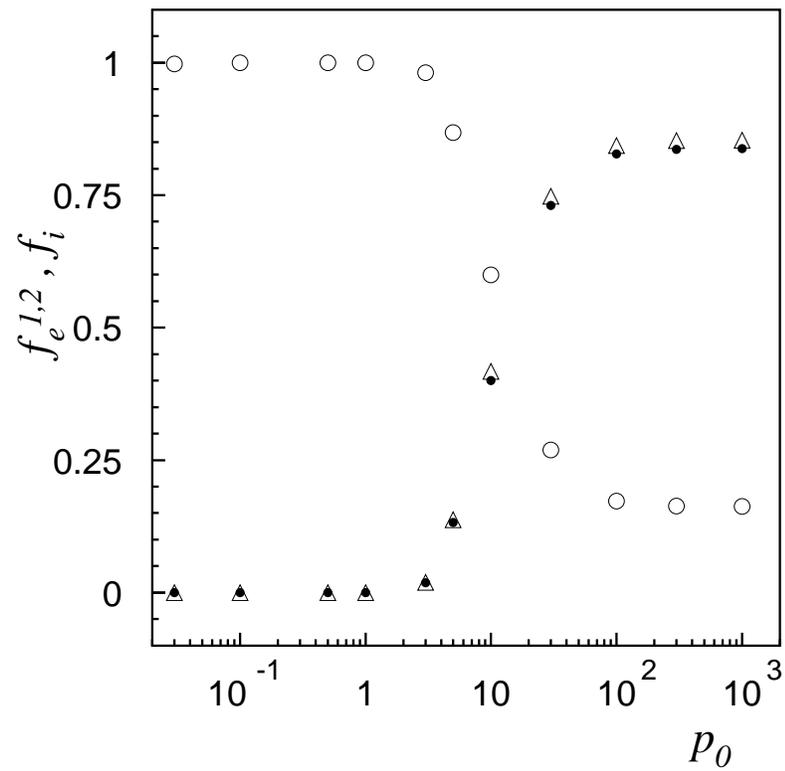}
\end{center}
\caption{Fractions $f_e^1$ ($\triangle$), $f_e^2$ ($\bullet$), and
$f_i$ ($\circ$) of the 10,000 configurations which contribute to
$N_e^1$, $N_e^2$, and $N_i$ in Fig. 4, as a function of $p_0$.}
\label{fig5}
\end{figure}

\end{document}